\documentclass[12pt]{article}

\usepackage{epsfig,amsmath,amssymb,graphics,graphicx,xcolor}

\numberwithin{equation}{section}

\newcommand{\rt}[1]{\textcolor{red}{#1}}
\newcommand{\bs}[1]{\boldsymbol{#1}}

\begin{document}
	
	
	\begin{center}
		
		
		\Large{\bf Fractional Generalizations of Gradient Mechanics }
   		    
   		 \vskip 7mm
   		 {\bf \large E. C. Aifantis} \\
   		 \vskip 3mm  
   		 
   		 
   		 \vskip 5mm
   		 {\normalsize Aristotle University of Thessaloniki, Thessaloniki, 
   		 54124, 
   		 Greece} \\
   
   		 {\normalsize Michigan Technological University, Houghton, MI 
   		 49931, USA} \\
   	 
   		 {\normalsize Togliatti State University, Togliatti 445020, Russia}
   		 
   		 \vskip 2mm
   		 
   		 {\small \textit{mom@mom.gen.auth.gr, ORCID: 0000-0002-6846-5686, 
   		 		Tel.:+30-2310995921}} \\
   	 		
   	 	\vskip 5mm 	
   	 	{\large \bf Abstract}
   		 
   \end{center}

    This short chapter provides a fractional generalization of gradient 
    mechanics, an approach (originally advanced by the author in the mid 80’s) 
    that has gained world-wide attention in the last decades due to its 
    capability of modeling pattern forming instabilities and size effects in 
    materials, as well as eliminating undesired elastic singularities. It is 
    based on the incorporation of higher-order gradients (in the form of 
    Laplacians) in the classical constitutive equations multiplied by 
    appropriate internal lengths accounting for the geometry/topology of 
    underlying micro/nano structures. This review will focus on the fractional 
    generalization of the gradient elasticity equations (GradEla) – an 
    extension of classical elasticity to incorporate the Laplacian of Hookean 
    stress – by replacing the standard Laplacian by its fractional counterpart. 
    On introducing the resulting fractional constitutive equation into the 
    classical static equilibrium equation for the stress, a fractional 
    differential equation is obtained whose fundamental solutions are derived 
    by using the Green’s function procedure. As an example, Kelvin’s problem is 
    analyzed within the aforementioned setting. Then, an extension to consider 
    constitutive equations for a restrictive class of nonlinear elastic 
    deformations and deformation theory of plasticity is pursued. Finally, the 
    methodology is applied for extending the author’s higher-order diffusion 
    theory from the integer to the fractional case.
	
	
	\section{Introduction}
	
	This contribution concerns the fractional generalization of the author’s 
	gradient elasticity and higher-order diffusion. Both of these theories were 
	introduced three decades ago to model deformation and transport problems in 
	media with micro/nanostructures. A new Laplacian term was added in the 
	standard constitutive equations of Hookean elasticity and Fickean diffusion 
	to interpret experimental data that could not be modeled by classical 
	theories. Among the new results were the elimination of undesirable elastic 
	singularities in dislocation lines and crack tips, as well as new robust 
	continuum models for grain boundary diffusion. All these problems were 
	successfully and efficiently addressed by incorporating internal lengths in 
	the standard constitutive equations of elasticity and diffusion, as scalar 
	multipliers of newly introduced Laplacian terms of the persistent 
	constitutive variables to account for “weakly” nonlocal effects. The 
	resulting internal length gradient (ILG) framework and its applications to 
	various areas of material mechanics are reviewed in a recent article by the 
	author \cite{ECA_ILG}, where extensive bibliography can also be found. In 
	the same article a brief account of fractional and fractal generalization 
	of the ILG framework is given. 
	
	We expand on the aforementioned review by providing an updated summary of 
	the fractional generalization of the ILG framework focusing on static 
	elasticity with a few related remarks on plasticity and steady-state 
	diffusion. In that connection, it is noted that the basic balance laws for 
	the mass and momentum are assumed to retain their classical (integer) form. 
	For stationary deformation and steady-state diffusion problems these laws 
	lead to the standard balance equations
	
	\begin{equation} \label{Equillibrium_Law}
		div \bs{\sigma} = 0 \quad or \quad \sigma_{ij,j} = 0 \,, 
	\end{equation}
	
	\noindent for the stress tensor and  
	
	\begin{equation}  \label{Conservation_Law}
		div \bs{j} = 0 \quad or \quad j_{i,i} = 0 \,,
	\end{equation}
	
	\noindent for the diffusive flux vector.
	
	The standard constitutive equations of the ILG framework for $\bs{\sigma}$  
	and $\bs{j}$ are of 
	the form
	
	\begin{align} \label{ConstEq_GradEla}
		{\bs \sigma} &= \lambda (tr{\bs \varepsilon}) \bs{1} +2\mu{\bs 
		\varepsilon}- 
		l_{\varepsilon}^2\nabla^2 [\lambda (tr{\bs \varepsilon}) \bs{1} 
		+2\mu{\bs \varepsilon}] ; \notag
		\\ 
		{\sigma_{ij}} &= \lambda \varepsilon_{kk} \delta_{ij} +2\mu 
		\varepsilon_{ij} -
		l_{\varepsilon}^2\nabla^2 [\lambda \varepsilon_{kk} \delta_{ij} +2\mu 
		\varepsilon_{ij}],
	\end{align}
	
	\noindent and 
	
	\begin{equation} \label{ConstEq_GradDiff}
	{\bs j} = -D \nabla ( \rho  - 
	l_{d}^2\nabla^2 \rho) ;  \quad 
    j_i = -D (\rho - l_{d}^2\nabla^2 \rho)_{\!,i} \,,
	\end{equation}
	
	\noindent respectively. The classical elastic moduli $(\lambda,\mu)$ and 
	diffusivity ($D$) have their usual meaning, the quantities 
	$\varepsilon_{ij}$ and $\rho$ denote strain and concentration respectively, 
	while the newly introduced parameters $l_{\varepsilon}$  and $l_d$ are 
	deformation--induced and diffusion-induced internal lengths (ILs) 
	accounting for “weakly” nonlocal effects. 
	
	The fractional generalization of the above equations consists of replacing 
	the standard (integer) Laplacian $\Delta$ in Eqs. \eqref{ConstEq_GradEla} 
	and \eqref{ConstEq_GradDiff} with a fractional one of the Riesz form 
	$(-{^R\Delta})^{\alpha/2} $ or the Caputo form $^C\Delta^\alpha_W$. 
	Then, 
	the corresponding fractional generalization of Eq. \eqref{ConstEq_GradEla} 
	reads
	
	\begin{equation} \label{ConstEq_GradEla_Riesz}
		{\sigma_{ij}} = (\lambda \varepsilon_{kk} \delta_{ij} +2\mu 
		\varepsilon_{ij}) -
		l_{\varepsilon}^2(\alpha) (-{^R\Delta})^{\alpha/2} [\lambda 
		\varepsilon_{kk} \delta_{ij} +2\mu 
		\varepsilon_{ij}],
	\end{equation}
	
	\noindent where $(-{^R\Delta})^{\alpha/2}$ is the fractional generalization 
	of the 
	Laplacian in the Riesz form, and 
	
	\begin{equation} \label{ConstEq_GradEla_Caputo}
		{\sigma_{ij}} = (\lambda \varepsilon_{kk} \delta_{ij} +2\mu 
		\varepsilon_{ij}) -
		l_{\varepsilon}^2(\alpha) ^C\Delta^\alpha_W [\lambda 
		\varepsilon_{kk} \delta_{ij} +2\mu 
		\varepsilon_{ij}],
	\end{equation}
	
	\noindent where $^C\Delta^\alpha_W$ is the fractional Laplacian in the 
	Caputo form 
	\cite{TarasovECA_NonlSciComm}. 
	
	Equations \eqref{ConstEq_GradEla_Riesz} and 
	\eqref{ConstEq_GradEla_Caputo} are fractional generalizations of the 
	original GradEla model.
	Similar equations can be written down for the fractional generalization of 
	Eq. \eqref{ConstEq_GradDiff}. They read
	
	\begin{equation} \label{ConstEq_GradDiff_Riesz}
		{\bs j} = -D \nabla [ \rho  - 
		l_{d}^2(\alpha) \{(-{^R\Delta})^{\alpha/2} \rho\}] ;  \quad 
		j_i = -D [\rho - l_{d}^2(\alpha) \{ (-{^R\Delta})^{\alpha/2} 
		\rho\}]_{\!,i} 
		\,,
	\end{equation}
	
	\noindent and 
	
	\begin{equation} \label{ConstEq_GradDiff_Caputo}
		{\bs j} = -D \nabla [ \rho  - 
		l_{d}^2(\alpha) \{^C\Delta^\alpha_W \rho\}] ;  \quad 
		j_i = -D [\rho - l_{d}^2(\alpha) \{ ^C\Delta^\alpha_W 
		\rho\}]_{\!,i} 
		\,,
	\end{equation}
	
	\noindent respectively.
	
	On introducing the aforementioned fractional gradient constitutive 
	equations into the non-fractional balance laws given by Eqs. 
	\eqref{Equillibrium_Law} and \eqref{Conservation_Law}, the corresponding 
	partial differential equations of fractional order 
	are obtained which need to be solved with the aid of appropriate boundary 
	conditions for finite domains. To dispense with the complication of 
	higher-order fractional boundary conditions, we consider infinite domains 
	and derive fundamental solutions for the respective problems by employing a 
	fractional extension of the Green’s function method.
	The above is illustrated in detail in the next section (Section 2) by 
	considering the classical Thomson (Lord Kelvin) elasticity problem in the 
	framework of fractional GradEla. In Section 3 we give a brief account on 
	preliminaries of fractional gradient nonlinear elasticity or deformation 
	theory of plasticity. We note that both of these sections are an update of 
	the fractional considerations of \cite{ECA_ILG} based on the detailed 
	elaborations contained in the initial articles by Tarasov and the 
	author  \cite{TarasovECA_JMBM},\cite{TarasovECA_NonlSciComm}, as well 
	as subsequent further discussions by Tarasov 
	\cite{Tarasov_Lattice1}--\cite{Tarasov_GradEla2}. In Section 4 we return 
	to the problem of fractional Laplacian and obtain fundamental solutions for 
	a general fractional equation of Helmholtz type which turns out to govern 
	both gradient elasticity and higher-order diffusion theory. Since the 
	basics of fractional deformation have been outlined in Sections 2 and 3, we 
	show in Section 4 how these basic results are directly applicable to 
	fractional diffusion problems. In this connection, it is noted that 
	solutions of fractional GradEla problems are reduced to solutions of an 
	inhomogenous Helmholtz equation, which is also the governing equation of 
	fractional generalization of electrostatics with Debye screening 
	\cite{Tarasov_Electrodynamics}.

	
	\section{Gradient Elasticity (GradEla): Revisiting Kelvin’s Problem}
	
	To shed light on the implications of fractional GradEla, a specific 3D 
	configuration with spherical symmetry is considered below. The model of Eq. 
	\eqref{ConstEq_GradEla_Riesz} is employed due to the fact that definite 
	results are available for the fractional Laplacian of Riesz type. The 
	corresponding most general fractional GradEla governing equation is of the 
	form \cite{TarasovECA_JMBM}  
	
	\begin{equation}  \label{Balance_Law_GradEla}  
		c_\alpha ( (-\Delta)^{\alpha/2}u)(\bs r)+ c_\beta 
		((-\Delta)^{\beta/2}u)(\bs{r}) = f(\bs{r})  \quad (\alpha > \beta),
	\end{equation}
	
	\noindent where $\bs{r} \in \mathbb{R}^3$ and $r = |\bs{r}|$  are 
	dimensionless,   $ (-\Delta)^{\alpha/2} $ is the Riesz fractional Laplacian 
	of 	order $\alpha$ , with the same for the symbols characterized by 
	$\beta$, and the (fractional) gradient coefficients $(c_\alpha,c_\beta)$ 
	are material constants related to the elastic moduli and the internal 
	length, respectively. The rest of the symbols have their usual meaning: $u$ 
	denotes displacement and $f(\bs{r}$ body force. For  $\alpha>0$ and 
	suitable functions $u(\bs{r}$ , the Riesz fractional 
	derivative can be defined in terms of the Fourier transform $\mathcal{F}$  
	by  
	
	\begin{equation}
		( (-\Delta)^{\alpha/2}u)(\bs r) = \mathcal{F}^{-1} (|\bs{k}|^\alpha 
		(\mathcal{F}u)(\bs{k})), 
	\end{equation}
	
	\noindent where $\bs{k}$ denotes the wave vector. If $\alpha = 4$ and 
	$\beta=2$, we have the well-known GradEla equation
	
	\begin{equation} \label{HighOrder_Balance_Law}
		c_2 \Delta u(\bs{r}) - c_4 \Delta^2 u(\bs{r}) + f(\bs{r}) = 0,
	\end{equation}
	
	\noindent where $c_2 = (\lambda + 2 \mu)$ and $c_4 = \pm (\lambda + 
	2\mu)\,l_s$ for 
	spherically symmetric problems. Equation \eqref{HighOrder_Balance_Law} is a 
	fractional partial differential equation with a solution of the form
	
	\begin{equation}
		u(\bs{r}) = \int_{\mathbb{R}^3} G_{\alpha,\beta}(\bs{r} - \bs{r}') 
		f(\bs{r}')\, d^3{\bs{r}'},
	\end{equation}
	
	\noindent with the Green-type function $G_{\alpha,\beta}$  given by 
	
	\begin{equation} \label{Greens_Function_Thompson}
		G_{\alpha,\beta}(\bs{r}) = \int_{\mathbb{R}^3} \frac{1}{
		c_\alpha |\bs{k}|^\alpha + c_\beta |\bs{k}|^\beta} 
		e^{i\bs{k}\cdot\bs{r}}
		\, d^3\bs{k} = 
		\frac{1}{(2\pi)^{3/2}\,\sqrt{|\bs{r}|}} \int_0^\infty 		
		\frac{\lambda^{3/2} J_{1/2}(\lambda |\bs{r}|) }{c_\alpha\lambda^\alpha 
		+ c_\beta \lambda^\beta}\, d\lambda, 
	\end{equation}
	
	\noindent where $J_{1/2}(z) = \sqrt{2/\pi z} \sin(z)$ is the Bessel 
	function of the first kind and the dot denotes inner product. 
	
	To proceed further, we consider Thomson's problem of an applied point load  
	$f_0$ , i.e. 
	
	\begin{equation}
		f(\bs{r}) = f_0 \delta(\bs{r}) = f_0 \delta(x) \delta(y) \delta(z).
	\end{equation}
	
	Then, the displacement field $u(\bs{r})$ has a simple form given by the 
	particular solution  
	
	\begin{equation}
		u(\bs{r}) = f_0 G_{\alpha,\beta} (\bs{r}), 
	\end{equation}	
	
	\noindent with  the Green's function given by Eq. 
	\eqref{Greens_Function_Thompson}, 
	i.e.  
	
	\begin{equation} \label{Greens_Function_Thompson_Exp}
		 u(\bs{r}) = \frac{f_0}{2 \pi^2 |\bs{r}|} \int_0^\infty 
		 \frac{\lambda \sin(\lambda|\bs{r}|)}{c_\alpha\lambda^\alpha 
		 	+ c_\beta \lambda^\beta} \, d\lambda, \quad (\alpha>\beta).
	\end{equation}
	
	It turns out that the asymptotic form of the solution given by Eq. 
	\eqref{Greens_Function_Thompson_Exp} for $0<\beta<2$, and $\alpha \ne 2$, 
	reads   
	
	\begin{equation}
		u(\bs{r}) \approx \frac{f_0 \Gamma(2-\beta) \sin(\pi\beta/2)}{2\pi^2 
		c_\beta} \frac{1}{|\bs{r}|^{3-\beta}}, \quad (|\bs{r}| \to \infty).
	\end{equation}
	
	This asymptotic behavior for $|\bs{r}| \to \infty$  does not depend on the 
	parameter $\alpha$, and (as will be seen below) the corresponding 
	asymptotic behavior for $|\bs{r}| \to 0$  does not 
	depend on the parameter  $\beta$ , where $\alpha>\beta$. It follows that 
	the displacement field at large distances from the point of load 
	application is determined only by the term $(-\Delta)^{\beta/2}$, where 
	$\beta<\alpha$. This can be interpreted as a fractional non-local 
	``deformation'' counterpart of the classical elasticity result based on 
	Hooke's law. We can also note the existence of a maximum for the quantity   
	$u(\bs{r})\cdot|\bs{r}|$ in the case $0<\beta<\alpha<2$ . Indeed, these 
	observations become clear by considering in detail the following two 
	special cases that emerge.
	
	\noindent {\bf A)} Sub-GradEla model: $\alpha = 2; 0 <\beta<2$. In this 
	case Eq. 
	\eqref{Balance_Law_GradEla} becomes  
	
	\begin{equation}  \label{Balance_Law_SubGradEla}  
		c_2 \Delta u(\bs r) - c_\beta 
		((-\Delta)^{\beta/2}u)(\bs{r}) + f(\bs{r}) = 0,  \quad (0 < \beta <2).
	\end{equation}
	
	The order of the fractional Laplacian $(-\Delta)^{\beta/2}$  is less than 
	the order of the first term related to the usual Hooke's law. For example, 
	one can consider the square of the Laplacian, i.e. $\beta=1$ . In general, 
	the parameter $\beta$  defines the order of the power-law non-locality. The 
	particular solution of Eq. \eqref{Balance_Law_SubGradEla} in the present 
	case, reads 
	
	\begin{equation} \label{Greens_Function_SubGradEla}
		u(\bs{r}) = \frac{f_0}{2 \pi^2 |\bs{r}|} \int_0^\infty 
		\frac{\lambda \sin(\lambda|\bs{r}|)}{c_2\lambda^2 
			+ c_\beta \lambda^\beta} \, d\lambda, \quad (0<\beta<2).
	\end{equation} 
	
	The following asymptotic behavior for Eq. 
	\eqref{Greens_Function_SubGradEla} can be derived in the form  
	
	\begin{equation} \label{Greens_Function_SubGradEla_Asympt}
		u(\bs{r}) = \frac{f_0}{2 \pi^2 |\bs{r}|} \int_0^\infty 
		\frac{\lambda \sin(\lambda|\bs{r}|)}{c_2\lambda^2 
		+ c_\beta \lambda^\beta} \approx
	    \frac{C_0(\beta)}{|\bs{r}|^{3-\beta}} + \sum_{k=1}^\infty
	    \frac{C_k(\beta)}{|\bs{r}|^{(2-\beta)(k+1)+1}} \quad (|\bs{r}| \to 
	    \infty), 
	\end{equation} 
	
	\noindent where 
	
	\begin{align}
		C_0(\beta) &=  \frac{f_0 \Gamma(2-\beta) \sin(\pi\beta/2)}{2\pi^2 
		c_\beta}, \notag \\
		C_k(\beta) &= -\frac{f_0 c_2^k}{2\pi^2c_\beta^{k+1}} \int_0^\infty
		z^{(2-\beta)(k+1)-1} \sin(z) \,dz.
	\end{align}
	
	As a result, the displacement field generated by the force that is applied 
	at a point in the fractional gradient elastic continuum described by the 
	fractional Laplacian $(-\Delta)^{\beta/2}$  with $0<\beta<2$ is given by
	
	\begin{equation}
		u(\bs{r}) \approx \frac{C_0(\beta)}{|\bs{r}|^{3-\beta}} \quad 
		(0<\beta<2),
	\end{equation}
	 
	\noindent for large distances $ (|\bs{r}| \to \infty)$. 
	
	\noindent {\bf B)} Super-GradEla model: $\alpha>2$  and $\beta=2$ . In this 
	case, 
	Eq. \eqref{Balance_Law_GradEla} becomes 
	
	\begin{equation}  \label{Balance_Law_SuperGradEla}  
		c_2 \Delta u(\bs r) - c_\alpha 
		((-\Delta)^{\alpha/2}u)(\bs{r}) + f(\bs{r}) = 0,  \quad (\alpha >2).
	\end{equation}
	
	The order of the fractional Laplacian $(-\Delta)^{\alpha/2}$  is greater 
	than the order of the 
	first term related to the usual Hooke's law. The parameter $\alpha>2$  
	defines the 
	order of the power-law non-locality of the elastic continuum. If $\alpha=4$ 
	, Eq. \eqref{Balance_Law_SuperGradEla} reduces to Eq. 
	\eqref{HighOrder_Balance_Law}. The case   can 
	be viewed as corresponding as 
	closely as possible ($\alpha\approx4$) to the usual gradient elasticity 
	model of Eq. \eqref{HighOrder_Balance_Law}. The asymptotic behavior of the 
	displacement field  $u(\bs{r})$ for $\bs{r}\to0$ in the 
	case of super-gradient elasticity is given by  
	
	\begin{align} \label{Greens_Function_SuperGradEla_Asympt}
		u(\bs{r}) &\approx 
		 \frac{f_0\Gamma((3-\alpha)/2)}{2^\alpha \pi^2 \sqrt{\pi} c_\alpha  
		 \Gamma(\alpha/2)} \frac{1}{|\bs{r}|^{3-\alpha}}, \quad (2<\alpha<3), 
		 \notag \\ 
	 	u(\bs{r}) &\approx 
	 	 \frac{f_0}{2\pi\alpha c_2^{1-3/\alpha} c_\alpha^{3/\alpha}
	 	 \sin(3\pi/\alpha)}, \quad\, (\alpha>3).
	\end{align}

	Note that the above asymptotic behavior does not depend on the parameter  
	$\beta$, and that the corresponding relation of Eq. 
	\eqref{Greens_Function_SuperGradEla_Asympt} does not depend on $c_\beta$ . 
	The displacement field $u(\bs{r})$ for short distances away from the point 
	of load application is determined only by the term with 
	$(-\Delta)^{\alpha/2},\, (\alpha>\beta)$, i.e. the fractional 
	counterpart of the usual extra non-Hookean term of gradient elasticity. 
	More details for the above results can be found in 
	\cite{TarasovECA_JMBM}--\cite{Tarasov_GradEla1}. (\rt{Shoud we add more?})
	
	
	\section{Fractional Gradient Plasticity}
	
	In this section we provide an introductory account of fractional 
	deformation theory of plasticity (as opposed ot the flow theory of 
	plasticity) by simply elaborating on a specific constitutive equation which 
	could also be viewed as a very special form of nonlinear elasticity. 
	
	The corresponding (nonlinear) fractional gradient constitutive equation 
	involves scalar measures of the stress and strain tensors; i.e. their 
	second invariants, as these quantities enter in both theories of nonlinear 
	elasticity and plasticity. In plasticity theory, in particular, we employ 
	the second invariants of the deviatoric stress and plastic strain tensors. 
	The effective (equivalent) stress $\sigma$ is defined by the equation
	
	\begin{equation}
		\sigma = \sqrt{(1/2) \sigma^{\prime}_{ij} \sigma^{\prime}_{ij}} \,; 
		\quad
		\sigma^{\prime}_{ij} = \sigma^{\prime}_{ij} 
		-\frac{1}{3}\sigma^{\prime}_{kk} \delta_{ij},
	\end{equation}
	
	\noindent where $\sigma_{ij}$ is the stress tensor. The effective 
	(equivalent) 
	plastic strain is defined as
	
	\begin{equation}
		\varepsilon = \int dt \sqrt{2 \dot{\varepsilon_{ij}} 
			\dot{\varepsilon_{ij}} },
	\end{equation}
	
	\noindent where $\dot{\varepsilon_{ij}}$ is the plastic strain rate tensor, 
	which is assumed to be traceless in order to satisfy plastic 
	incompressibility.
	
	Motivated by the above, we propose the following form of nonlinear 
	fractional differential equation for the scalar quantities $\sigma$ and 
	$\varepsilon$ which 
	can be used as a basis for future tensorial formulation of nonlinear 
	elasticity and plasticity theories
	
	\begin{equation} \label{Nonlinear_ConstEq}
		\sigma({\bf r}) =
		E \, \varepsilon ({\bf r}) + c (\alpha)
		\, ((-\Delta)^{\alpha/2} \varepsilon ) ({\bf r}) +
		\eta \, K(\varepsilon ({\bf r})), \quad (\alpha >0),
	\end{equation}
	
	\noindent where $K(\varepsilon ({\bf r}))$ is a nonlinear function,
	which describes the usual (homogeneous) material's response (linear 
	hardening plasticity); $c (\alpha)$ is an internal parameter,
	that measures the nonlocal character of deformation mechanisms; 
	$E$ is a shear-like elastic modulus; $\eta$ is a small parameter of 
	nonlinearity; and$(-\Delta)^{\alpha/2}$ is the fractional Laplacian
	in the Riesz form.
	As a simple example of the nonlinear function, we can consider 
	
	\begin{equation} \label{Nonlinear_Function}
		K(\varepsilon ) = \varepsilon^{\beta} ({\bf r}) ,
		\quad (\beta >0) .
	\end{equation} 
	
	Equation \eqref{Nonlinear_ConstEq}, where $K(\varepsilon(\bs{r}))$ is 
	defined by Eq. \eqref{Nonlinear_Function} is the fractional 
	Ginzburg--Landau equation. 
	For various choices of the parameters $(E,\eta,\beta)$ characterizing the 
	homogenous material response, different models of nonlinear elastic and 
	plastic behavior may result. 
	
	Let us derive a particular solution of Eq. \eqref{Nonlinear_ConstEq}  
	with  $K(\varepsilon ({\bf r}))=0$. To solve the 
	linear fractional differential equation 
	
	\begin{equation} \label{Linear_ConstEq}
		\sigma({\bf r}) = E \, \varepsilon ({\bf r}) +
		c (\alpha) \, [(-\Delta)^{\alpha/2} \varepsilon ] ({\bf r}) ,
	\end{equation}
	
	we apply the Fourier method and the fractional Green function method.
	Using Theorem 5.22 of Kilbas et al  (\cite{Kilbas_FDE} pages 342 and 343, 
	see also \cite{Samko_Kilbas}) for the case $E \ne 0$ and $\alpha > 
	(n-1)/2$, we see that Eq. \eqref{Linear_ConstEq} is solvable, 
	and its particular solution is given by the equation 
	
	\begin{equation} \label{Conv_ConsEq_Lin}
		\varepsilon ({\bf r}) = G_{n,\alpha}  *  \sigma = 
		\int_{\mathbb{R}^n} G_{n,\alpha} ({\bf r} - {\bf r}^{\prime}) \, 
		\sigma ({\bf r}^{\prime}) \, d^n {\bf r}^{\prime} ,
	\end{equation}
	
	\noindent where the symbol $*$ denotes convolution, and
	$G_{n,\alpha} ({\bf r})$ is defined by 
	
	\begin{equation} \label{Greens_Function_ConsEq_Lin}
		G_{n,\alpha} ({\bf r}) = \frac{|{\bf r}|^{(2-n)/2}}{(2 \pi)^{n/2}} 
		\int^{\infty}_0 
		\frac{ \lambda^{n/2} \, J_{(n-2)/2} (\lambda |{\bf r}|) }{c (\alpha) \, 
		\lambda^{\alpha} +E} \, d \lambda,
	\end{equation}
	
	\noindent where $n=1,2,3$, $\alpha > (n-1)/2$, and 
	$J_{(n-2)/2}$ is the Bessel function of the first kind.
	
	Let us consider a deformation of unbounded fractional nonlocal continuum, 
	where the stress is applied to an infinitesimally small region in this 
	continuum. In this case, we can assume that the strain 
	$\varepsilon(\bs{r})$ is induced by a point stress $\sigma(\bs{r})$  at the 
	origin of coordinates, i.e. 
	
	\begin{equation} \label{Point_Stress}
		\sigma({\bf r}) = \sigma_0 \delta({\bf r}),
	\end{equation}
	
	\noindent i.e. the particular solution is proportional to the Green's 
	function. As a 
	result, the stress field is
	
	\begin{equation} \label{Greens_Function_ConsEq_Lin_Strain}
		\varepsilon ({\bf r}) = 
		\frac{1}{2 \pi^2} \frac{\sigma_0}{|{\bf r}|} \, 
		\int^{\infty}_0 \frac{ \lambda \, \sin (\lambda |{\bf r}|)}{ 
			E + c (\alpha) \, \lambda^{\alpha} } \, d \lambda .
	\end{equation}
	
	\subsection{Perturbation of Linearized Fractional Deformations by Nonlinear 
	Hardening}

	Suppose that $\varepsilon(\bs{r}) = \varepsilon_0(\bs{r})$ is the solution 
	of Eq. \eqref{Nonlinear_ConstEq} with $\eta=0$, i.e.  is the 
	solution of the linear equation 
	
	\begin{equation}
		\sigma({\bf x}) =
		E \, \varepsilon_0 ({\bf r}) +
		c (\alpha) \, ((-\Delta)^{\alpha/2} \varepsilon_0) ({\bf r}) 
	\end{equation}
	
	The solution of this fractional differential equation may be written in the 
	form 
	
	\begin{equation}
		\varepsilon ({\bf r}) = \varepsilon_0 ({\bf r}) + 
		\eta \, \varepsilon_1 ({\bf r}) + \ldots
	\end{equation}
	
	This means that we consider perturbations to the strain field 
	$\varepsilon_0(\bs{r})$ of the fractional gradient deformation state, which 
	are caused by weak nonlinear hardening effects. The first order 
	approximation with respect to $\eta$ gives 
	the equation 
	
	\begin{equation} \label{Linear_ConstEq_Eff}
		E \, \varepsilon_1 ({\bf r}) 
		+ c (\alpha) \, ((-\Delta)^{\alpha/2} \varepsilon_1) ({\bf r}) 
		+ K (\varepsilon_0({\bf r})) = 0 ,
	\end{equation}
	
	\noindent which is equivalent to the linear equation 
	
	\begin{equation}
		\sigma_{eff}({\bf r}) = E \, \varepsilon_1 ({\bf r}) +
		c (\alpha) \, ((-\Delta)^{\alpha/2} \varepsilon_1) ({\bf r}) ,
	\end{equation}
	
	\noindent where the effective stress $\sigma_{eff} ({\bf x})$
	is defined by the equation
	
	\begin{equation} \label{Sigma_Eff}
		\sigma_{eff} ({\bf r}) = - K (\varepsilon_0({\bf r})) .
	\end{equation}
	Equation \eqref{Linear_ConstEq_Eff} gives a particular solution in the form
	
	\begin{equation} \label{Epsilon_Pert}
		\varepsilon ({\bf r}) = \varepsilon_0 ({\bf r}) + \varepsilon_1 ({\bf 
		r}) =
		G_{n,\alpha}  *  \sigma + \eta \, G_{n,\alpha} *  
		\sigma_{eff} ,
	\end{equation} 
	
	\noindent where the convolution operation and $G_{n,\alpha}$ are defined by 
	Eqs. \eqref{Conv_ConsEq_Lin}, \eqref{Greens_Function_ConsEq_Lin}. Upon 
	substitution of Eq. \eqref{Sigma_Eff}  into Eq. \eqref{Epsilon_Pert}, we 
	obtain 
	
	\begin{equation} \label{Epsilon_Pert2}
		\varepsilon ({\bf r}) = 
		G_{n,\alpha} *  \sigma - \eta \, G_{n,\alpha}  *  K( 
		G_{n,\alpha}  *  \sigma ) . 
	\end{equation}
	
	For a ``point stress'' of the form given by Eq. \eqref{Point_Stress}, Eq. 
	\eqref{Epsilon_Pert} can be 
	written in the form 
	
	\begin{equation}
		\varepsilon ({\bf r}) = 
		\sigma_0 \, G_{n,\alpha}({\bf r}) - \eta \, \Bigl( G_{n,\alpha} \, * \, 
		K( \sigma_0 \, G_{n,\alpha} ) \Bigr)({\bf r}) . 
	\end{equation}
	
	\noindent which, for  $K(\cdot)$ given by Eq. \eqref{Nonlinear_Function}, 
	results to
	
	\begin{equation}
		\varepsilon ({\bf r}) = 
		\sigma_0 \, G_{n,\alpha}({\bf r}) - \eta \, \sigma^{\beta}_0 \, 
		\Bigl( G_{n,\alpha} \, * \, (G_{n,\alpha})^{\beta} \Bigr) ({\bf r}).
	\end{equation}
	
	\subsection{Perturbation of Plasticity by Fractional Gradient Nonlocality}
	
	Let us now consider an equilibrium state by setting $\varepsilon_0 
	=const.$  (i.e. $(-\Delta)^{\alpha/2}\varepsilon_0 = 0 $ ) and 
	$\sigma(\bs{r}) = \sigma = const.$  in Eq. \eqref{Nonlinear_ConstEq}, i.e.
	
	\begin{equation} \label{Nonlinear_AlgEq}
		E \, \varepsilon_0 + \eta \, K(\varepsilon_0) = \sigma .
	\end{equation}
	
	This, for the case, where the function $K$ is defined by Eq. 
	\eqref{Nonlinear_Function} with $\beta=3$, 
	becomes
	
	\begin{equation}
		E \, \varepsilon_0 + \eta \, \varepsilon ^3_0 = \sigma . 
	\end{equation}
	
	For $\sigma \ne 0$, there is no solution $\varepsilon_0=0$. 
	For $E >0$ and the weak scalar stress field $\sigma\ll \sigma_c$ 
	with respect to the critical value $\sigma_c = \sqrt{E / \eta}$,  there 
	exists only one solution
	
	\begin{equation}
	\varepsilon_0 \approx \sigma / E .
	\end{equation}
	
	For negative stiffness materials ($E <0$) and $\sigma=0$, 
	we have three solutions
	
	\begin{equation}
		\varepsilon_0 \approx \pm \sqrt{|E| / \eta } , 
		\quad \varepsilon_0 =0 .
	\end{equation}
	
	For $\sigma < (2 \sqrt{3}/ 9) \sigma_c$, 
	also exist three solutions. 
	For $\sigma \gg \sigma_c$, 
	we can neglect the first term ($E \approx 0$),
	
	\begin{equation}
		\eta \, \varepsilon ^3_0 \approx \sigma ,
	\end{equation}
	
	\noindent and obtain 
	
	\begin{equation}
		\varepsilon_0 \approx (\sigma / \eta)^{1/3} . 
	\end{equation}
	
	In general, the equilibrium values  $\varepsilon_0$  are solutions of the 
	nonlinear 
	algebraic relation given by Eq. \eqref{Nonlinear_AlgEq}. 

	Let us consider a deviation $\varepsilon_1(\bs{r})$ of the 
	field from the equilibrium value $\varepsilon_0(\bs{r})$. For this purpose 
	we will seek a solution in the form
	
	\begin{equation}
		\varepsilon ({\bf r})= \varepsilon_0 + \varepsilon_1 ({\bf r})
	\end{equation}
	
	In general, the stress field is not constant, i.e. 
	$\sigma({\bf x}) \ne \sigma$. In a first  approximation, we 
	obtain the equation 
	
	\begin{equation}  \label{Linearized_ConstEq}
		\sigma({\bf r}) =
		c (\alpha) \, ((-\Delta)^{\alpha/2} \varepsilon_1) ({\bf r}) 
		+ \Bigl( E + \eta \, K^{\prime}_{\varepsilon }(\varepsilon_0) \Bigr) 
		\varepsilon_1 ({\bf r}) ,
	\end{equation}
	
	\noindent where $K^{\prime}_{\varepsilon } = \partial K(\varepsilon ) / 
	\partial \varepsilon $.Equation \eqref{Linearized_ConstEq} is equivalent to 
	the linear fractional differential equation
	
	\begin{equation}
		\sigma({\bf x}) =
		E_{eff} \, \varepsilon_1 ({\bf r}) +
		c (\alpha) \, ((-\Delta)^{\alpha/2} \varepsilon_1) ({\bf r}), 
	\end{equation}
	
	\noindent with the effective modulus $E_{eff}$ defined by 
	
	\begin{equation}
		E_{eff} = E + \eta \, K^{\prime}_{\varepsilon }(\varepsilon_0) .
	\end{equation}
	
	For the case $K(\varepsilon )=\varepsilon^{\beta}$, we have 
	
	\begin{equation}
		E_{eff} = E + \beta \, \eta \, \varepsilon ^{\beta-1}_0 .
	\end{equation}
	
	A particular solution of Eq. \eqref{Linearized_ConstEq} can be written in 
	the form of Eq. \eqref{Conv_ConsEq_Lin}, where we use $E_{eff}$ instead of  
	$E$ For the ``point stress'' (see Eqs. 
	\eqref{Point_Stress}--\eqref{Greens_Function_ConsEq_Lin_Strain}), Eq. 
	\eqref{Epsilon_Pert2} gives
	
	\begin{equation}
		\varepsilon_1 ({\bf r}) = 
		\frac{1}{2 \pi^2} \frac{\sigma_0}{|{\bf r}|} \, 
		\int^{\infty}_0 
		\frac{E + E_{eff} + 2 c (\alpha) \, \lambda^{\alpha}}{ 
			( c (\alpha) \, \lambda^{\alpha}+ E ) \,
			( c (\alpha) \, \lambda^{\alpha}+ E_{eff} ) } 
		\, \sin (\lambda |{\bf r}|) \, d \lambda .
	\end{equation}
	
	For the case $\alpha=2$, 
	the field $\varepsilon_1({\bf r})$ is given by
	the equation 
	
	\begin{equation}
		\varepsilon_1 ({\bf r}) = \frac{\sigma_0}{4 \pi c (\alpha) \, |{\bf 
		r}|} \, e^{ - |{\bf r}|/ r_c } ,
	\end{equation}
	
	\noindent where $r_c$ is defined by 
	
	\begin{equation}
		r^2_c = \frac{c (\alpha)}{ E + \eta \, K^{\prime}_{ \varepsilon 
		}(\varepsilon_0)} .
	\end{equation}
	
	It should be noted that on analogous situation exists
	in classical theory of electric fields.
	In the electrodynamics the field $\varepsilon_1({\bf r})$ describes 
	the Coulomb potential with the Debye screening.
    For a fractional differential field equation  ($\alpha \ne 2$), 
	we have a power-law type of screening 
	that is described in \cite{Tarasov_Electrodynamics}.
	The electrostatic potential 
	for media with power-law spatial dispersion
	differs from the Coulomb potential by the factor
	
	\begin{equation}
		C_{\alpha,0} (|{\bf r}|) =
		\frac{2}{\pi} \, 
		\int^{\infty}_0 \frac{ \lambda \, \sin (\lambda |{\bf r}|)}{
			E_{eff} + c (\alpha) \, \lambda^{\alpha} } \, d \lambda.
	\end{equation}
	
	Note that the Debye potential differs from the Coulomb potential by the 
	exponential factor 
	$C_D(|{\bf x}|) =\exp (-|{\bf r}|/r_D)$. 
	
	
	\section{Fractional Helmholtz Equation}
	
	On introducing the fractional GradEla constitutive relation given by Eq. 
	\eqref{ConstEq_GradEla_Riesz} into the equilibrium relation given by Eq. 
	\eqref{Equillibrium_Law}, we obtain 
	
	\begin{equation} \label{Equil_GradEla1}
	[ 1+ l_\varepsilon^\alpha(-\Delta)^{\alpha/2}][\lambda \nabla 
	tr\bs{\varepsilon}+2\mu\, 
	div\bs{\varepsilon}] = 0, 
	\end{equation}
	
	\noindent where the notation $l_\varepsilon^2(\alpha) \equiv 
	l_\varepsilon^\alpha $, and $  
	(-{^R\Delta})^{\alpha/2} \equiv (-\Delta)^{\alpha/2}$ was used for 
	simplicity. Noting the fact that the operators $\nabla$ and $ 
	(-\Delta)^{\alpha/2}$  commute and that the 
	second bracket in Eq. \eqref{Equil_GradEla1} is also zero by replacing  
	$\bs{\varepsilon}$ with $\bs{\varepsilon}_0$, where $\bs{\varepsilon}_0$  
	denotes the solution of the corresponding equation for classical 
	elasticity, (i.e. $\lambda \nabla 
	tr\bs{\varepsilon}+2\mu\, div\bs{\varepsilon}=0$ ), we can easily deduce 
	that the solution of Eq. \eqref{Equil_GradEla1} satisfies the reduced 
	fractional partial 
	differential equation 
	
	\begin{equation} \label{Ru_Aifantis_GradEla}
		[ 1+ l_\varepsilon^\alpha(-\Delta)^{\alpha/2}] \bs{\varepsilon} = 
		\bs{\varepsilon}_0,
	\end{equation}
	
	\noindent which for the case $\alpha = 2$ reduces to the inhomogeneous 
	Helmholtz equation derived for the non-fractional GradEla (Ru-Aifantis 
	theorem \cite{Ru_Aifantis}) and was 
	used successfully to derive non-singular solutions for dislocations and 
	cracks \cite{Gutkin_ECA}--\cite{ECA_TAMS}. It turns out that compatible 
	displacements $\bs{u}$ ($ \varepsilon_{ij} = (1/2)[u_{i,j}+u_{j,i}] $) also 
	obey Eq.\eqref{Ru_Aifantis_GradEla} and the same holds for corresponding 
	fields in electrostatics 
	with Debye screening \cite{Tarasov_Electrodynamics}, as well as for 
	steady-state higher-order 
	diffusion problems \cite{ECA_GradNano}, \cite{ECA_DiffSolids}. 
	
	It is thus critical to derive fundamental solutions for 
	Eq.\eqref{Ru_Aifantis_GradEla}; i.e. for the equation
	
	\begin{equation} \label{Helmholtz_Equation}
	[ 1+ l_\varepsilon^\alpha(-\Delta)^{\alpha/2}] G_\alpha(\bs{r}) = 
	\delta(\bs{r}),
	\end{equation}
	
	\noindent where $G_\alpha(\bs{r})$ denotes the fundamental solution, 
	$\delta(\bs{r})$ denotes the delta function and $\bs{r}$  is the radial 
	coordinate in a 3D space.
	
	To compute the fundamental solution of Eq. \eqref{Helmholtz_Equation} with 
	the natural boundary 
	condition $G_\alpha(\bs{r}) \to 0$  as $\bs{r}\to\infty$, we employ the 
	method of Fourier transforms. Using the properties of the Fourier transform 
	of the Riesz fractional Laplacian for 
	every ``well-behaved'' scalar function $f(\bs{r})$ 
	
	\begin{equation} \label{Fourier_Riesz_Laplacian}
		\mathcal{F} 
		(\,(-\Delta)^{\alpha/2}f(\bs{r}) \,) (\bs{k}) = 
		\left|\bs{k}\right|^\alpha 
		\mathcal{F} (f(\bs{r}))(\bs{k}) \,,  
	\end{equation}
	
	\noindent and the well-known transform of the delta function $\mathcal{F} 
	(\delta(\bs{r}))(\bs{k}) = 1 $, we obtain the following algebraic equation
	for the fundamental solution
	
	\begin{equation} \label{Helmholtz_Fractional_Fourier1}
		[\, 1+l_\varepsilon^{\,\alpha} \left|\bs{k}\right|^{\alpha}] \, 
		G_\alpha(\bs{k}) = 1, 
	\end{equation} 
	
	\noindent which gives
	
	\begin{equation} \label{Helmholtz_Fractional_Fourier2}
		G_\alpha(\bs{k}) =  \frac{1}{ 1+l_\varepsilon^{\alpha} 
		\left|\bs{k}\right|^{\alpha}}
	\end{equation}  
	
	Consequently, the fundamental solution of Eq. \eqref{Helmholtz_Equation} in 
	the physical space 
	is obtained through inversion of Eq. \eqref{Helmholtz_Fractional_Fourier2}
	
	\begin{equation}  \label{Helmholtz_Fractional_Fourier_Inversion}
		G_\alpha(\bs{r}) = \frac{1}{(2\pi)^3} \int_{-\infty}^{\infty} \, 
		\frac{1}{ 1+l_\varepsilon^{\alpha} \left|\bs{k}\right|^{\alpha}} \, 
		e^{i 
		\bs{k}\cdot\bs{r}} \, d^3\bs{k}.
	\end{equation}
	
	To simplify Eq. \eqref{Helmholtz_Fractional_Fourier_Inversion}, we perform 
	a change of variables $\bs{k}\rightarrow l_\varepsilon^{-1} \bs{k}$, which 
	results a factor of $l_\varepsilon^{-3}$  and a change in scale 
	$\bs{r}\rightarrow\bs{r}/l_\varepsilon$. Therefore, for simplicity, we omit 
	those factors, and restore them at the end result. 
	
	The integral given by Eq. \eqref{Helmholtz_Fractional_Fourier_Inversion} is 
	defined in a 3-dimensional Euclidean space and can be analytically computed 
	in spherical coordinates by applying a well-known relationship (see, for 
	example Lemma 25.1 of Samko et al \cite{Samko_Kilbas})  
	
	\begin{equation}\label{LemmaSamkoKilbasElasticity} 
	\frac{1}{(2\pi)^{3}}
	\int_{-\infty}^{\infty} f(\left|\bs{k}\right|) e^{i\bs{k}\cdot\bs{r}} 
	\, d^3\bs{k} = 
	\frac{1}{(2\pi)^{3/2}\sqrt{|\bs{r}|}}
	\int_{0}^{\infty} k^{3/2}\,f(k) \,J_{1/2}(k |\bs{r}|) \, dk.
	\end{equation} 
	
	In Eq. \eqref{LemmaSamkoKilbasElasticity} $k$ denotes the magnitude of the 
	wave vector and $J_{1/2} = \sqrt{2/(\pi z)} \sin(z)$ denotes the Bessel 
	function of order $1/2$. Introduction of Eq.  
	\eqref{LemmaSamkoKilbasElasticity} in Eq. 
	\eqref{Helmholtz_Fractional_Fourier_Inversion} by omitting 
	the scaling factors, results to 
	
	\begin{align} \label{RadialIntegralElasticity}
	G_\alpha (\bs{r}) &= 
	\frac{1}{(2\pi)^{3/2}\sqrt{|\bs{r}|}}
	\int_{0}^{\infty} \frac{k^{3/2}}{1+k^\alpha} \,J_{1/2}(k |\bs{r}|) \, dk, 
	\notag \\ 
	& = \frac{1}{2\pi^2|\bs{r}|} \int_0^\infty \frac{k}{1+k^\alpha} 
	\sin(k|\bs{r}|)\, dk.
	\end{align}
	
	The integral in Eq. \eqref{RadialIntegralElasticity} can be computed using 
	the convolution property of the 
	Mellin transform, defined in \cite{Erdelyi_IT} by the relationship 
	
	\begin{equation} \label{Mellin_Transform}
		\mathcal{M}(f(x)) (s) = \int_{0}^{\infty} f(x) \, x^{s-1} \,dx.
	\end{equation}	 
	
	\noindent Its inverse is given by 
	
	\begin{equation} \label{Inverse_Mellin_Transform}
		f(x)  = \frac{1}{2\pi i} \int_{\gamma-i\infty}^{\gamma+i\infty} f(s) 
		\, x^{-s} \,ds, 
	\end{equation}	 
	
	where the path of integration is a vertical strip separating the poles 
	of 	$\mathcal{M}(f(x)) (s) $, defined in $ \gamma_1 < Re(s) < \gamma_2$. 
	For more details about the Mellin transform, we refer the reader to 
	\cite{Marichev_IT}. Here we only use the basic results 
	
	\begin{align}
	\mathcal{M}(\frac{1}{1+x^\alpha})(s) &=  \frac{1}{\alpha} \,
	\Gamma(\frac{s}{\alpha})  \, \Gamma(1-\frac{s}{\alpha}), \notag \\
	\mathcal{M}(x^{3/2}J_{1/2}(x))(s) &= \, 2^{1/2 \, + s } \, 
	\frac{\Gamma\left(1+\frac{s}{2}\right)}
	{\Gamma\left(\frac{1}{2}-\frac{s}{2}\right)},
	\end{align}  
	
	\noindent where we made use of the Mellin transform of the Bessel function 
	(see also Section 6.8 of \cite{Erdelyi_IT}) 
	
	\begin{equation}
	\mathcal{M} (J_\sigma(2\sqrt{u}))(s) = 
	\frac{\Gamma\left(\frac{\sigma}{2}+s\right)}
	{\Gamma\left(\frac{\sigma}{2}+1-s\right)}.
	\end{equation} 
	
	Consequently, Eq. \eqref{RadialIntegralElasticity} can be evaluated using 
	the above results and 
	performing the inverse Mellin transform by computing the Mellin-Barnes 
	integral 
	
	\begin{equation}	\label{MellinBarnes_GradEla1}	
	G_\alpha(\bs{r}) = \frac{1}{2 \alpha 
		\pi^{3/2}\, |\bs{r}|^{2}} \, 
	\frac{1}{2\pi 
		i } \int_{\gamma-i\infty}^{\gamma+i\infty}
	\frac{ \Gamma(\frac{1}{\alpha}-\frac{s}{\alpha}) \, 
	\Gamma(1-\frac{1}{\alpha}+\frac{s}{\alpha})
		\Gamma\left(1+\frac{s}{2}\right)}
	{\Gamma\left(\frac{1}{2}-\frac{s}{2}\right)}
	\, \left(\frac{|\bs{r}|}{2}\right)^{-s} \, ds.
	\end{equation}
	
	The Mellin-Barnes integral representation of Eq. 
	\eqref{MellinBarnes_GradEla1} can be expressed in 
	terms of the corresponding Fox-H function of fractional analysis (see, for 
	example, \cite{Metzler_Klafter}--\cite{Luchko})
	
	\begin{equation}  \label{H_Function_GradEla1}
	G_\alpha(\bs{r}) = \frac{1}{2 \alpha 
		\pi^{3/2}\, |\bs{r}|^{2}} \,
	H^{2,1}_{1,3} \left[ 
	\frac{|\bs{r}|}{2} \,\,\bigg| 
	\arraycolsep=1.pt\def\arraystretch{1.2} 
	\begin{array}{ll}
	\left(1-\frac{1}{\alpha},\frac{1}{\alpha}\right) \\
	\left(1-\frac{1}{\alpha},\frac{1}{\alpha}\right),
	\left(1,\frac{1}{2}\right),
	\left(\frac{1}{2},\frac{1}{2}\right)
	\end{array}
	\right].
	\end{equation}
	
	The integral \eqref{MellinBarnes_GradEla1} has poles at the points 
	$ s = 1 -\alpha (\nu -1) $ and $ s = 1-(3+2\nu) $, $\nu \in \mathbb{N}$. To 
	evaluate it we apply first the residue theorem to the poles of f $ 
	\Gamma(1-\frac{1-s}{\alpha})$, since they 
	correspond to the singularity near the origin $r\approx 0$.
	
	After a change of variables $ s\rightarrow s + 1-\alpha $, 
	\eqref{MellinBarnes_GradEla1} becomes \cite{Mathai}--\cite{Luchko}
	
	\begin{equation}	\label{MellinBarnesInt2Elasticity}	
	G_\alpha(\bs{r}) = \frac{1}{ \alpha \,2^\alpha\,  
		\pi^{3/2}\, |\bs{r}|^{3-\alpha}} \, 
	\frac{1}{2\pi i} \int_{\gamma-i\infty}^{\gamma+i\infty}
	\frac{ \Gamma(1-\frac{s}{\alpha}) \, \Gamma(\frac{s}{\alpha})
		\Gamma\left(\frac{3}{2} -\frac{\alpha}{2} + 
		\frac{s}{2}\right)}{\Gamma\left(\frac{\alpha}{2} - \frac{s}{2}\right)}
	\, \left(\frac{|\bs{r}|}{2}\right)^{-s} \, ds.
	\end{equation}
	
	Next, we perform another change of variables $s\rightarrow\alpha s$ in Eq. 
	\eqref{MellinBarnesInt2Elasticity}, which results 
	to 
	
	\begin{equation}	\label{MellinBarnesInt3Elasticity}	
	G_\alpha(\bs{r}) = \frac{1}{\,2^\alpha\,  
		\pi^{3/2}\, |\bs{r}|^{3-\alpha}} \, 
	\frac{1}{2\pi i} \int_{\gamma-i\infty}^{\gamma+i\infty}
	\frac{ \Gamma(1-s) \, \Gamma(s)
		\Gamma\left(\frac{3}{2} -\frac{\alpha}{2}(1 - s)\right)}
	{\Gamma\left(\frac{\alpha}{2}(1-s) \right)} \,
	\left(\frac{|\bs{r}|}{2}\right)^{-\alpha s} \, ds,
	\end{equation}
	
	\noindent which provides an alternative representation in terms 
	of the Fox-H function, i.e. 
	
	\begin{equation}  \label{H_Function_GradEla2}
	G_\alpha(\bs{r}) = \frac{1}{\,2^\alpha\,  
		\pi^{3/2}\, |\bs{r}|^{3-\alpha}}  \,
	H^{2,1}_{1,3} \left[ 
	\left(\frac{|\bs{r}|}{2}\right)^\alpha \,\,\bigg| 
	\arraycolsep=1.pt\def\arraystretch{1.2} 
	\begin{array}{ll}
	\left(0,1\right) \\
	\left(0,1\right),
	\left(\frac{3}{2}-\frac{\alpha}{2},\frac{\alpha}{2}\right),
	\left(1-\frac{\alpha}{2},\frac{\alpha}{2}\right)
	\end{array}
	\right].
	\end{equation}
	
	The contour integral in Eq. \eqref{H_Function_GradEla2} can be evaluated 
	using the method of residues from complex analysis, by closing the contour 
	encircling all poles at $s=-\nu$ and then applying the Cauchy residue 
	theorem  
	
	\begin{equation} \label{ResiduesGradEla}
	G_\alpha(\bs{r}) = \frac{1}{\,2^\alpha\,  
		\pi^{3/2}\, |\bs{r}|^{3-\alpha}} 
	\sum_{\nu=0}^{\infty} \lim_{s\to-\nu} \, \left\{
	(s+\nu) \Gamma(s) \, \frac{\Gamma(1-s)  \,
		\Gamma\left(\frac{3}{2} -\frac{\alpha}{2}(1 - s)\right)}
	{\Gamma\left(\frac{\alpha}{2}(1-s) \right)} \,
	\left(\frac{|\bs{r}|}{2}\right)^{-\alpha s} \right\}.
	\end{equation} 
	
	Eq. \eqref{ResiduesGradEla} can be evaluated using the relation  
	
	\begin{equation} \label{ResidueSimpleElasticity}
	\lim_{s\to-\nu} 
	(s+\nu) \, \Gamma(s) = \lim_{s\to-\nu} 
	\frac{\Gamma(s+\nu+1)}{s(s+1)..(s+\nu-1)} = \frac{(-1)^\nu }{\nu\,!}.
	\end{equation}
	
	This gives 
	
	\begin{equation} \label{SeriesExpansion1Elasticity}
	G_\alpha(\bs{r}) = \frac{1}{\,2^\alpha\,  
		\pi^{3/2}\, |\bs{r}|^{3-\alpha}} 
	\sum_{\nu=0}^{\infty} \frac{(-1)^\nu }{\nu\,!} \, 
	\frac{\Gamma(1+\nu)  \,
		\Gamma\left(\frac{3}{2} -\frac{\alpha}{2}(1 +\nu)\right)}
	{\Gamma\left(\frac{\alpha}{2}(1+\nu) \right)} \,
	\left(\frac{|\bs{r}|}{2}\right)^{\alpha \nu}, 
	\end{equation} 
	
	\noindent which can be simplified by noting 
	that $\Gamma(1+\nu)=\nu! $, for $\nu\in\mathbb{N}$. 
	The final result is 
	
	\begin{equation} \label{SeriesExpansion2Elasticity}
	G_\alpha(\bs{r}) = \frac{1}{\,2^\alpha\,  
		\pi^{3/2}\, |\bs{r}|^{3-\alpha}} 
	\sum_{\nu=0}^{\infty}  \,
	\frac{\Gamma\left(\frac{3}{2} -\frac{\alpha}{2}(1 +\nu)\right)}
	{\Gamma\left(\frac{\alpha}{2}(1+\nu) \right)} \,
	(-1)^\nu
	\left(\frac{|\bs{r}|}{2}\right)^{\alpha \nu}. 
	\end{equation}
	
	An asymptotic expression near the origin is obtained from the dominating 
	term of Eq. \eqref{SeriesExpansion2Elasticity} for $r\to0$, i.e. 
	
	\begin{equation} \label{AsymptoticGradEla}
	G_\alpha(\bs{r}) \approx \frac{\Gamma\left(\frac{3}{2} 
	-\frac{\alpha}{2}\right)}{\,2^\alpha\,  
		\pi^{3/2}\Gamma(\frac{\alpha}{2})} 
	 \, \frac{1}{|\bs{r}|^{3-\alpha}}, \quad (\bs{r} \to 0).
	\end{equation}. 
	
	This asymptotic form cancels the singularity of the fundamental solution of 
	corresponding classical theories. To see this, one can compute the 
	contributions from the poles of $\Gamma(1+s/2)$  in Eq. (4.14), which 
	correspond to 
	non-singular asymptotic behavior near the origin, using the same 
	techniques. The result is
	
	\begin{align} \label{SeriesExpansion3Elasticity}
		G_\alpha(\bs{r}) &= \frac{1}{\,2^\alpha\,  
			\pi^{3/2}\, |\bs{r}|^{3-\alpha}} 
			\sum_{\nu=0}^{\infty}  \,
			\frac{\Gamma\left(\frac{3}{2} -\frac{\alpha}{2}(1 +\nu)\right)}
			{\Gamma\left(\frac{\alpha}{2}(1+\nu) \right)} \,
			(-1)^\nu
			\left(\frac{|\bs{r}|}{2}\right)^{\alpha \nu} \notag\\
			&+ \frac{2}{\alpha (4\pi)^{3/2}} \sum_{\nu=0}^\infty
			\frac{\Gamma\left(\frac{3+2\nu}{\alpha}\right) 
			      \Gamma\left(1-\frac{1}{\alpha}(3+2\nu) \right)}
			{\Gamma\left(\frac{3}{2}+\nu \right)} \frac{(-1)^\nu}{\nu!}
			\left(\frac{|\bs{r}|}{2}\right)^{2\nu}.
	\end{align}
	
	In the special case $\alpha \to 2$, Eq. \eqref{SeriesExpansion3Elasticity} 
	reduces to the Green's function of the classical Helmholtz equation, i.e.
	
	\begin{equation}
		G_\alpha(\bs{r}) = \frac{1}{4\pi |\bs{r}|}e^{-|\bs{r}|}
	\end{equation}
	
	It is easily checked that Eqs. \eqref{AsymptoticGradEla}, 
	\eqref{SeriesExpansion3Elasticity}  give the same results as Eq. (64) of 
	\cite{Tarasov_Electrodynamics}, since it solves the same mathematical 
	equation (i.e. the fractional inhomogeneous Helmholtz equation (their Eq. 
	(59)) for a different physical problem -- the problem of a point charge.
	
	
	\section{Fractional Higher-Order Diffusion}
	
	On introducing the fractional diffusion constitutive relation given by Eq. 
	\eqref{ConstEq_GradDiff_Riesz} into the classical (non-fractional) mass 
	balance law 
	
	\begin{equation} \label{Mass_Balance_Law}
		\frac{\partial\rho}{\partial t} + div \bs{j} = 0 \quad or \quad
		\rho_{,t} + j_{i,i} = 0,
	\end{equation}
	
	\noindent we obtain the fractional high-order diffusion equation
	
	\begin{equation} \label{High_Order_Diff_Eq}
		\frac{\partial\rho}{\partial t} = D \Delta \rho + D l_d^\alpha 
		\nabla\cdot \left\{(-\Delta)^{\alpha/2} \nabla\rho \right\},
	\end{equation}
	
	\noindent along with the auxiliary conditions 
	$\rho(\bs{r},0)=\delta(\bs{r}),\,\,\rho(\bs{r},t)\to0 \,\, as\,\, 
	|\bs{r}|\to\infty$  and   $\delta(\bs{r})$ denoting, as usual, the delta 
	function. [The notation $l_d^2(\alpha)\equiv l_d^\alpha$, and 
	$(-{^R\Delta})^{\alpha/2} \equiv (-\Delta)^{\alpha/2}$  was used 
	for simplicity]
	
	To solve Eq. \eqref{High_Order_Diff_Eq} we employ the method of Fourier 
	transform and exploit 
	the properties of the Riesz fractional Laplacian, along with the symmetry 
	of the problem. This gives 
	
	\begin{equation} \label{Ordinary_Diffrential_Equation_Fourier}
		\frac{\partial \rho(\bs{k},t)}{\partial t} = -D \left| 
		\bs{k}\right|^2 \rho(\bs{k},t) - D \, l_d^{\alpha}  
		\left| \bs{k}\right|^{\alpha} 
		\left| \bs{k}\right|^2  \rho(\bs{k},t), 
	\end{equation}     
	
	\noindent where  $\bs{k}$ denotes the wave vector. Equation 
	\eqref{Ordinary_Diffrential_Equation_Fourier} is a first order 
	ordinary differential equation with respect to time with the initial 
	condition $\rho(\bs{k},0) = \mathcal{F}(\delta(\bs{r})) = 1$. Its solution 
	is    
	
	\begin{equation} \label{Ordinary_Diffrential_Equation_Solution}
		\rho(\bs{k},t)= \exp( -D t \left| \bs{k}\right|^2\,)  \,
		\exp( -D_\alpha t \left| \bs{k}\right|^{\alpha+2}\,),
	\end{equation}
	
	\noindent where we defined $D_\alpha \equiv D l_d^\alpha$. The solution of 
	Eq. \eqref{Ordinary_Diffrential_Equation_Solution}
	in configuration space is 
	obtained by inversion of the Fourier transform 
	
	\begin{equation} \label{Inversion_Fourier_Transform}
		\rho(\bs{r},t)  = \frac{1}{(2\pi)^{3}}\int_{-\infty}^{\infty} 
		\exp( -D t \left| \bs{k}\right|^2\,)  
		\exp( -D_\alpha t \left| \bs{k}\right|^{\alpha+2}\,)  
		\exp{(i \bs{k}\cdot \bs{r})} \,d^3\bs{k}.
	\end{equation}
	
	Equation \eqref{Inversion_Fourier_Transform} is the inverse Fourier 
	transform of the product of two 
	independent terms and can be expressed as the convolution of the 
	corresponding solutions in the physical space using the following 
	well-known property of the Fourier transform
	
	\begin{align} \label{Fourier_Conv_Diff}
	\mathcal{F} ( (f\ast g) (\bs{r},t) ) (\bs{k}) &= 
	\mathcal{F} ( f(\bs{r},t) ) (\bs{k}) \,
	\mathcal{F} ( g(\bs{r},t) ) (\bs{k}), \\
	( f\ast g) (\bs{r},t)  &= \int_{-\infty}^{\infty} 
	f(\bs{r}-\bs{r}^\prime,t) g(\bs{r}^\prime,t) \, d^3\bs{r}^\prime.
	\end{align}
	
	Using Eq. \eqref{Fourier_Conv_Diff}, we recognize Eq. 
	\eqref{Inversion_Fourier_Transform} as 
	the convolution   
	
	\begin{equation} \label{Convolution_Solution_Fractional_Diffusion}
		\rho(\bs{r},t) = ( G_{2} \ast  G_{\alpha+2} \,)  (\bs{r},t), 
	\end{equation}  
	
	\noindent where we defined the set of functions $G_\alpha$ as
	
	\begin{align}  \notag
	G_{\alpha} (\bs{r},t) &= \mathcal{F}^{-1} \{ 
	\exp(-D_\alpha t \left|\bs{k}\right|^\alpha  ) \}  \\
	&= \frac{1}{(2\pi)^3} \int_{-\infty}^{\infty}
	\exp(-D_\alpha t \left|\bs{k}\right|^\alpha  ) \exp(i\bs{k}\cdot\bs{r})
	\, d^3\bs{k}  \label{Fundamental_Solution_Diffusion_Fourier}.
	\end{align}   
	
	Equation \eqref{Fundamental_Solution_Diffusion_Fourier} is the fundamental 
	solution 
	(i.e. the Green’s function) for the fractional diffusion equation 
	
	\begin{equation} 
	\frac{\partial G_\alpha(\bs{r},t)}{\partial t} =  
	-D_\alpha  \left(-\Delta\right)^{\alpha/2} 
	G_\alpha(\bs{r},t), \label{Fundamental_Diffusion_Equation}
	\end{equation}
	
	The corresponding fundamental solution of Eq. \eqref{High_Order_Diff_Eq} is 
	then deduced from 	Eq. \eqref{Fundamental_Solution_Diffusion_Fourier} 
	through the convolution property of Eq. 
	\eqref{Convolution_Solution_Fractional_Diffusion}
	
	Applying  Eq. \eqref{LemmaSamkoKilbasElasticity} into the fundamental 
	solution $G_\alpha(\bs{r})$  of Eq. 
	\eqref{Fundamental_Solution_Diffusion_Fourier}, we obtain 
	
	\begin{align}
		G_{\alpha}(\bs{r},t) &= \frac{1}{(2\pi)^{3/2}\sqrt{|\bs{r}|}}
		\int_{0}^{\infty} k^{3/2}\,\exp(-D_\alpha t \, k^\alpha )
		\,J_{1/2}(k |\bs{r}|) \, dk \notag \\ 
		&=\frac{1}{2\pi^2|\bs{r}|} \int_{0}^{\infty} k\,\exp(-D_\alpha t 
		\, k^\alpha ) \sin(k|\bs{r}|)\,dk. \label{RadialIntegral} 
	\end{align} 
	
	The integral in Eq. \eqref{RadialIntegral} can be computed using the 
	convolution property of the Mellin transform, as in the previous section. 
	The final result is the following series expansion expression 
	\cite{ParisisECA_Diff}
	
	\begin{equation} \label{SeriesExpansionDiffusion}
		G_\alpha(\bs{r},t) = \frac{2}{\alpha 
		(4\pi)^{3/2} \,(D_\alpha\,t)^{3/\alpha}} 
		\sum_{\nu=0}^{\infty} \frac{(-1)^\nu}{\nu\,!} 
		\frac{ \Gamma(\frac{3}{\alpha} +\frac{2\nu}{\alpha}) }
		{\Gamma\left(\frac{3}{2} + \nu \right)} \,  
		\left(\!\frac{|\bs{r}|^2}{4(D_\alpha\, 
			t)^{2/\alpha}}\!\right)^{\nu} .
	\end{equation}
	
	Equation \eqref{SeriesExpansionDiffusion} can be represented in terms of 
	the Wright’s function $_1\Psi_1$ as 
	
	\begin{equation} \label{SeriesExpansionWright}
	G_\alpha(\bs{r},t) = \frac{2}{\alpha 
		(4\pi)^{3/2} \,(D_\alpha\,t)^{3/\alpha}} \,\,  _1\Psi_1 
	\left[ \arraycolsep=1.pt\def\arraystretch{1.2} 
	\begin{array}{l}
	\left(\frac{3}{\alpha},\frac{2}{\alpha}\right) \\
	\left(\frac{3}{2},1\right) 
	\end{array} 
	;\, -\frac{|\bs{r}|^2}{4 (D_\alpha \, t)^{\frac{2}{\alpha}}}
	\right].
	\end{equation}
	
	The generalized Wright’s function  is defined by the following series 
	\cite{Kilbas_FDE}, \cite{Samko_Kilbas}
	
	\begin{equation} \label{GeneralizedWright}
		_p\Psi_q (z) =  \, _{\!p}\Psi_q
		\left[  \arraycolsep=1.pt\def\arraystretch{1.2} 
		\begin{array}{lll}
		\left(a_1,A_1\right) & \ldots & \left(a_p,A_p\right)  \\
		\left(b_1,B_1\right) & \ldots & \left(b_q,B_q\right) 
		\end{array} 
		;\, z \right] = 
		\sum_{\nu=0}^{\infty} 
		\frac{\prod_{j=1}^{p} \Gamma(a_j+A_j\nu)}
		{\prod_{j=1}^{q} \Gamma(b_j+B_j\nu)} \,
		\frac{z^\nu}{{\nu\,!} }. 
	\end{equation}
	
	It is easy to check that when $\alpha=2$, the series expansion reduces to 
	the Green’s function of the ordinary diffusion equation in 3-dimensional 
	space. 	This is readily seen by letting $\alpha\to2$ in 
	Eqs. \eqref{SeriesExpansionDiffusion} and \eqref{SeriesExpansionWright}, 
	resulting to the expression 
	
	\begin{align} 
	G_2(\bs{r},t) &= \frac{1}{ 
		(4\pi D \,t)^{3/2}} \,\,  _1\Psi_1 
	\left[ \arraycolsep=1.pt\def\arraystretch{1.2} 
	\begin{array}{l}
	\left(\frac{3}{2},1\right) \notag \\
	\left(\frac{3}{2},1\right) 
	\end{array} 
	;\, -\frac{|\bs{r}|^2}{4 D \, t}
	\right] \notag \\  
	&= \frac{1}{ (4\pi D \,t)^{3/2}}	
	\sum_{\nu=0}^{\infty} \frac{(-1)^\nu}{\nu\,!} \, 
	\left(\frac{|\bs{r}|^2}{4 D \, t}\right)^\nu \notag \\ 
	\label{StandardDiffusion}
	& = \frac{1}{(4\pi D \,t)^{3/2}} \, \exp(-\frac{|\bs{r}|^2}{4 D \, t}).
	\end{align}
	
	Consequently, the fundamental solution of the second-order 
	fractional 	diffusion equation \eqref{High_Order_Diff_Eq}, denoted as 
	$G(\bs{r},t)$, is obtained through convolution of Eq. 
	\eqref{Fourier_Conv_Diff}, with   $G_\alpha(\bs{r},t)$
	given by Eq. \eqref{SeriesExpansionWright}	and \eqref{StandardDiffusion}
	for $\alpha=2$, i.e.
	
	\begin{equation} \label{Greens_Function_High_Ord_Diff}
		G(\bs{r},t) = \int_{-\infty}^\infty 
		G_{\alpha+2}(\bs{r}-\bs{r}^\prime,t) 
		G_2(\bs{r}^\prime,t)\,d^3\bs{r}^\prime. 
	\end{equation}
	
	We can extend Eq. \eqref{Mass_Balance_Law} to include distributed sources 
	(e.g. chemical reaction or trapping) with density/concentration rate 
	$q(\bs{r},t$. In this particular case, the classical mass balance law 
	becomes
	
	\begin{equation} \label{Mass_Balance_Sources}
	\frac{\partial\rho}{\partial t} + div \bs{j} = q, 
	\end{equation}
	
	\noindent and the corresponding inhomogeneous fractional diffusion equation 
	reads  
	
	\begin{equation} \label{High_Order_Diff_Eq_Sources}
		\frac{\partial\rho}{\partial t} = D \Delta \rho + D l_d^\alpha 
		\nabla\cdot \left\{(-\Delta)^{\alpha/2}\nabla \rho \right\} + q.
	\end{equation}
	
	Using the Fourier transform method, we can obtain the fundamental solution 
	of Eq. \eqref{High_Order_Diff_Eq_Sources} as follows 
	
	\begin{equation} \label{Greens_Function_High_Order_Sources}
		\rho(\bs{r},t) = \int_0^t\int_{-\infty}^{\infty} 
		G(\bs{r}-\bs{r}^\prime,t-\tau) 
		q(\bs{r}^\prime,\tau)\,d^3\bs{r}^\prime\,d\tau, 
	\end{equation}
	
	\noindent where $G(\bs{r},t)$  is given by Eq. 
	\eqref{Greens_Function_High_Ord_Diff}. 
	For the special case of a unit point source $ q(\bs{r},t) = 
	\delta(\bs{r})\delta(t)$, it is readily seen that Eq. 
	\eqref{Greens_Function_High_Order_Sources} reduces to the fundamental 
	solution $G(\bs{r},t)$.

	The fractional diffusion equation admits steady-state solutions, under the 
	presence of external sources/sinks with density/rate $q(\bs{r})$. The 
	governing equation for this time independent configuration is 
	
	\begin{equation} \label{High_Order_Steady_Sources}
	D \Delta \rho + D l_d^\alpha 
	\nabla\cdot \left\{(-\Delta)^{\alpha/2} \nabla\rho \right\} + q = 0.
	\end{equation}
	
	Equation \eqref{High_Order_Steady_Sources} can be generalized to a 
	higher-order steady-state fractional diffusion equation of the form
	
	\begin{equation}  \label{High_Order_GradDiff}  
	D_\alpha ( (-\Delta)^{\alpha/2}\rho)(\bs r)+ D_\beta 
	((-\Delta)^{\beta/2}\rho)(\bs{r}) = q(\bs{r}),  \quad (\alpha > \beta),
	\end{equation}
	
	\noindent where $(\alpha,\beta)$ denote arbitrary positive fractional order 
	and $(D_\alpha,D_\beta)$ are corresponding 
	fractional diffusion coefficients. Equation \eqref{High_Order_GradDiff}can 
	be derived by considering a fractional extension of the conservation law 
	given by Eq. \eqref{Mass_Balance_Sources}, along with the constitutive 
	relation given by Eq. \eqref{ConstEq_GradDiff_Riesz} and/or a 
	further fractional extension for its classical gradient ($\nabla$) part.
	
	Equation \eqref{High_Order_GradDiff} is a fractional partial differential 
	equation, whose solution reads
	
	\begin{equation} \label{Cov_Solution_Source}
		\rho(\bs{r}) = \int_{\mathbb{R}^3} G_{\alpha,\beta}(\bs{r} - \bs{r}') 
		q(\bs{r}')\, d^3{\bs{r}'},
	\end{equation}
	
	\noindent with the Green-type function $G_{\alpha,\beta}(\bs{r})$  given by 
	
	\begin{equation} \label{Greens_Function_Point_Source_Diff}
		G_{\alpha,\beta}(\bs{r}) = \int_{\mathbb{R}^3} \frac{1}{
		D_\alpha |\bs{k}|^\alpha + D_\beta |\bs{k}|^\beta} 
		e^{i\bs{k}\cdot\bs{r}}
		\, d^3\bs{k} = 
		\frac{1}{(2\pi)^{3/2}\,\sqrt{|\bs{r}|}} \int_0^\infty 		
		\frac{\lambda^{3/2} J_{1/2}(\lambda |\bs{r}|) }{D_\alpha\lambda^\alpha 
		+ D_\beta \lambda^\beta}\, d\lambda. 
	\end{equation}
	
	Let us now consider the particular problem of a unit point source located 
	at the origin of the form 
	
	\begin{equation} \label{Point_Source_Diff}
		q(\bs{r}) = q_0 \delta(\bs{r})  = q_0 \delta(x) \delta(y) \delta(z).
	\end{equation}
	
	Upon substitution of Eq. \eqref{Point_Source_Diff} into Eq. 
	\eqref{Cov_Solution_Source}, we 
	obtain the particular solution
	
	\begin{equation}
		\rho(\bs{r}) = q_0 G_{\alpha,\beta} (\bs{r}), 
	\end{equation}
	
	\noindent with the Green function $G_{\alpha,\beta} (\bs{r})$ given by Eq. 
	\eqref{Greens_Function_Point_Source_Diff}. By using then the 
	particular expression for the Bessel function of the first kind, we obtain
	
	\begin{equation} \label{Greens_Function_HigDiff_Source}
		\rho(\bs{r}) = \frac{q_0}{2 \pi^2 |\bs{r}|} \int_0^\infty 
		\frac{\lambda \sin(\lambda|\bs{r}|)}{c_\alpha\lambda^\alpha 
		+ c_\beta \lambda^\beta} \, d\lambda, \quad (\alpha>\beta).
	\end{equation}
	
	Two distinct modes of diffusion arise, depending on the particular form of 
	the fractional parameters $(\alpha,\beta)$, which are discussed in detail 
	below
	
	
	\noindent {\bf A)} Sub-GradDiffusion model: $\alpha = 2; 0 <\beta<2$. In 
	this case Eq. 
	\eqref{High_Order_Steady_Sources} becomes  
	
	\begin{equation}  \label{High_Order_SubDiff}  
	D \Delta \rho(\bs r) - D_\beta 
	((-\Delta)^{\beta/2}\rho)(\bs{r}) + q(\bs{r}) = 0,  \quad (0 < \beta 
	<2).
	\end{equation}
	
	The order of the fractional Laplacian $(-\Delta)^{\beta/2}$  is less than 
	the order of the first term related to the usual Fick's law. The parameter 
	$\beta$  defines the order of the power-law non-locality. The 
	particular solution of Eq. \eqref{High_Order_SubDiff} reads 
	
	\begin{equation} \label{Greens_Function_SubGradDiff}
	\rho(\bs{r}) = \frac{q_0}{2 \pi^2 |\bs{r}|} \int_0^\infty 
	\frac{\lambda \sin(\lambda|\bs{r}|)}{D\lambda^2 
		+ D_\beta \lambda^\beta} \, d\lambda, \quad (0<\beta<2).
	\end{equation} 
	
	The following asymptotic behavior for Eq. 
	\eqref{Greens_Function_SubGradDiff} can be derived in the form  
	
	\begin{equation} \label{Greens_Function_SubGradDiff_Asympt}
	\rho(\bs{r}) = \frac{q_0}{2 \pi^2 |\bs{r}|} \int_0^\infty 
	\frac{\lambda \sin(\lambda|\bs{r}|)}{D\lambda^2 
		+ D_\beta \lambda^\beta} \approx
	\frac{C_0(\beta)}{|\bs{r}|^{3-\beta}} + \sum_{k=1}^\infty
	\frac{C_k(\beta)}{|\bs{r}|^{(2-\beta)(k+1)+1}} \quad (|\bs{r}| \to 
	\infty), 
	\end{equation} 
	
	\noindent where 
	
	\begin{align}
	C_0(\beta) &=  \frac{q_0 \Gamma(2-\beta) \sin(\pi\beta/2)}{2\pi^2 
		D_\beta}, \notag \\
	C_k(\beta) &= -\frac{q_0 D^k}{2\pi^2D_\beta^{k+1}} \int_0^\infty
	z^{(2-\beta)(k+1)-1} \sin(z) \,dz.
	\end{align}
	
	As a result, the density of the diffusive species generated by the source 
	that is concentrated at a single point in space, for large distances from 
	the source, is given asymptotically by the expression
	
	\begin{equation}
	\rho(\bs{r}) \approx \frac{C_0(\beta)}{|\bs{r}|^{3-\beta}} \quad 
	(0<\beta<2),
	\end{equation}
	
	\noindent for large distances $ (|\bs{r}| \to \infty)$. 
	
	\noindent {\bf B)} Super-GradDiffusion: $\alpha>2$  and $\beta=2$. In this 
	case, Eq. \eqref{High_Order_Steady_Sources} becomes 
	
	\begin{equation}  \label{High_Order_SuperDiff}  
	D \Delta \rho(\bs r) - D_\alpha 
	((-\Delta)^{\alpha/2}\rho)(\bs{r}) + q(\bs{r}) = 0,  \quad (\alpha >2).
	\end{equation}
	
	The order of the fractional Laplacian $(-\Delta)^{\alpha/2}$  is greater 
	than the order of the 
	first term related to the usual Fick's law. The asymptotic density 
	$\rho(\bs{r})$ for $\bs{r}\to0$ in this case is given by  
	
	\begin{align} \label{Greens_Function_SuperGradDiff_Asympt}
	\rho(\bs{r}) &\approx 
	\frac{q_0\Gamma((3-\alpha)/2)}{2^\alpha \pi^2 \sqrt{\pi} D_\alpha 
		\Gamma(\alpha/2)} \frac{1}{|\bs{r}|^{3-\alpha}}, \quad (2<\alpha<3), 
	\notag \\ 
		\rho(\bs{r}) &\approx 
		\frac{q_0}{2\pi\alpha D^{1-3/\alpha} D_\alpha^{3/\alpha}
		\sin(3\pi/\alpha)}, \quad\, (\alpha>3).
	\end{align}
	
	Note that the above asymptotic behavior does not depend on the parameter  
	$\beta$, and that the corresponding relation of Eq. 
	\eqref{Greens_Function_SuperGradDiff_Asympt} does not depend on $D_\beta$ . 
	The density $\rho(\bs{r})$ for short distances away from the point 
	of source application is determined only by the term with 
	$(-\Delta)^{\alpha/2},\, (\alpha>\beta)$.
	
	Finally, and especially for the case of more complicated boundary value 
	problems, we mention that the steady--state diffusion of Eq. 
	\eqref{High_Order_Steady_Sources} can be 
	factored as 
	
	\begin{equation}
		D \nabla\cdot\nabla\left\{1+l_d^\alpha(-\Delta)^{\alpha/2} \right\} \rho
		+ q = 0.
	\end{equation}
	
	By defining the ``classical'' operator $L^0 \equiv D \nabla\cdot\nabla$, 
	and similarly its ``fractional gradient'' counterpart $L^\alpha \equiv 
	1+l_d^\alpha(-\Delta)$, we can prove that $L^\alpha$  satisfies the 
	classical steady-state Fickean diffusion equation. This is a direct 
	consequence of the fact that the operators $L^0$  and $L^\alpha$ commute. 
	Therefore, we arrive at the following ``operator-split'' scheme   
	
	\begin{equation} \label{Fractional_Ru_Aifantis_Diff}
		\left(1+l_d^\alpha(-\Delta)^{\alpha/2} \right) \rho = \rho_0;
		\quad  D \nabla\cdot\nabla\rho_0 + q = 0.
	\end{equation}    
	
	Equation \eqref{Fractional_Ru_Aifantis_Diff} is the fractional counterpart 
	of the Ru-Aifantis theorem \cite{Ru_Aifantis}, for the steady-state 
	fractional higher--order diffusion equation. 
	
	\begin{center}
		\section*{Acknowledgment}
	\end{center}
	
	\noindent Support of the Ministry of Education and Science of Russian 
	Federation under grant no. 14.Z50.31.0039 is acknowledged.


\end{document}